# Inefficiency of orbital Hall effect on the spin torque in transition metal/ferromagnet bilayers

Yizhuo Song, Jialin Tian, Fanxing Zheng, Jianting Dong, Meng Zhu, and Jia Zhang*

*School of Physics and Wuhan National High Magnetic Field Center,*

*Huazhong University of Science and Technology, 430074 Wuhan, China.*

*jiazhang@hust.edu.cn

## Abstract

Current induced spin torque is essential and crucial in spintronics. In this work, we systematically investigate the spin torque in transition metal(TM)/ferromagnet(FM) bilayers by using first-principles calculations and taking into account the phonon scattering at room temperature. To examine the spin and orbital Hall contribution, the studied transition metals include 5*d* heavy metals Pt, W, Au as well as 3*d* light metals Ti, V, Cr, Cu *etc*. We found that in TM/CoFe bilayers with typical 3*d* and 5*d* transition metals, the spin torque on CoFe mainly originates from spin Hall mechanism with the magnitude and sign of damping like torque efficiency consistent with the spin Hall conductivity. In TM/Ni bilayers, the spin torque is contributed by three mechanisms including spin and orbital Hall current in TM, as well as self-torque in Ni. The orbital Hall contribution in TM is accompanied by noteworthy opposite self spin torque in Ni, which leads to inapparent torque efficiency in Ti/Ni and V/Ni bilayers. For TM(5*d* heavy metal)/Ni bilayers, the spin torque induced by orbital Hall and self-torque in Ni nearly cancel each other, which makes the spin torque on Ni still align with that of the spin Hall effect in TM. Our work reveals much less efficient contribution of orbital Hall than spin Hall effect on the spin torque in transition metal/ferromagnet bilayers.

## Introduction

The spin torque (ST), which arises from the exchange interaction between electric field induced non-equilibrium spin density with the magnetization of magnetic materials, is a fundamental phenomenon in spintronics[1]. It plays a pivotal role in advanced magnetic memory technologies, particularly in magnetic random-access memory (MRAM). In the case of spin-orbit torque (SOT) in "heavy metal/ferromagnet" bilayers, two spin-orbit coupling driven mechanisms are involved, including the spin Hall effect (SHE) in bulk heavy metal[2][3] and the interfacial Rashba-Edelstein effect[4][5][6]. The spin current generated from SHE in nonmagnetic materials flows perpendicularly into the adjacent magnetic layer, is absorbed by the FM and leads to torques on magnetization. The Rashba-Edelstein effect prompts the generation of non-equilibrium spin polarization due to spin-momentum locking when an electric current flows along an interface with broken inversion symmetry.

Recently the orbital Hall effect (OHE), the counter part of SHE in bulk nonmagnetic metal (NM) has attracted intensive research attention since it can emerge even in light metals with negligible spin-orbit coupling (SOC), and with the characteristic orbital Hall conductivity (OHC) being much larger than the spin Hall conductivity (SHC)[7][8][9]. In NM/FM bilayers, when the resulting orbital Hall current in NM is injected into the adjacent FM layer, it may be converted into non-equilibrium spin density through SOC in the FM layer and exert an orbit torque (OT) on the FM layer[10][11][12]. It is thought that in particular cases the orbital torque (OT) may be comparable or even dominating over the SOT due to the large orbital Hall conductivity in NM, which may offer an alternative pathway for achieving highly efficient spin torque (ST) in spintronic devices.

In NM/FM bilayers, the spin and orbital Hall effects coexist and it is not clear and controversialto what extent the orbital Hall effect may contribute to the total spin torque. Therefore, it is urgently desirable to conduct comprehensive first-principles calculations to examine the main spin torque mechanism in NM/FM bilayers. In this work, we perform first-principles calculations on the spin torque in TM/CoFe and TM/Ni bilayers by choosing TM layers to be 5*d* heavy metals as well as 3*d* light metals. In TM/CoFe bilayers, the damping like torque has been solely contributed by spin Hall effect, whereas the signature of orbital Hall contribution in TM layer is absent. In TM/Ni bilayers, the spin torque can be attributed to the spin Hall and orbital Hall current in TM, as well as self-torque in Ni. The orbital Hall contribution is always accompanied by noteworthy self spin torque in Ni, which leads to small net damping like torque efficiency in Ti/Ni and V/Ni bilayers. For TM(5*d* heavy metal)/Ni bilayers, the resultant spin torque also aligns with the spin Hall effect in TM.

## Computational method

The first-principles calculations are performed by using fully relativistic multiple scattering Korringa-Kohn-Rostoker Green's function method[13][14]. For a magnetic system, the Dirac equation in density functional theory can be written as[13]:

$$[c\vec{\boldsymbol{\alpha}}\cdot\hat{\boldsymbol{p}}+(\beta-1)mc^2+V_{eff}(\boldsymbol{r})+\beta\vec{\Sigma}\cdot\vec{\boldsymbol{B}}_{xc}]\psi(\boldsymbol{r})=E\psi(\boldsymbol{r})$$

where $\vec{\boldsymbol{B}}_{xc}$ is the spin-dependent exchange-correlation energy (or effective magnetic field). When electric filed is applied to the NM/FM bilayers, the non-equilibrium spin-density $\delta\vec{\boldsymbol{S}}(\boldsymbol{r})$ will be induced, and the spin torque on magnetization at point $\boldsymbol{r}$ will be given by the direct exchange interaction between $\delta\vec{\boldsymbol{S}}(\boldsymbol{r})$ and $\vec{\boldsymbol{B}}_{xc}$ [15]:

$$\vec{\boldsymbol{T}}(\boldsymbol{r})=[\vec{\boldsymbol{B}}_{xc}(\boldsymbol{r})\times\delta\vec{\boldsymbol{S}}(\boldsymbol{r})]$$

By integrating over the magnetic atoms, it yields the total spin torque on magnetic layers:

$$\vec{\boldsymbol{T}}=\iiint_{\Omega_m}\vec{\boldsymbol{T}}(\boldsymbol{r})d\boldsymbol{r}$$

From the above general spin torque expressions, it is obvious that the SOT from spin related mechanisms *i.e.* spin Hall and Rashba-Edelstein effects are direct, while the orbital mechanism is indirect or second order effect. Since unlike spin Hall and spin current, the orbit degree of freedom does not directly interact with local magnetization. Instead, it needs to be converted into spin density *via* SOC in FM.

Under the linear response theory, the electric field induced torque $\vec{\boldsymbol{T}}$ on FM layer can be further simplified and calculated by first-principles calculations[16][17][18]:

$$\boldsymbol{T}_i = \boldsymbol{t}_{ij}\boldsymbol{E}_j \quad i,j \in x,y,z$$

where $t_{ij}$ is the electric field induced torkance tensor, $E_j$ is the electric field along $\boldsymbol{j}$ direction. The torque (torkance) can be classified as filed like (FL) and damping like (DL) terms, which are time reversal odd (T-odd) and even (T-even) respectively, *i.e* $\boldsymbol{T}_{FL}(-\boldsymbol{M})=-\boldsymbol{T}_{FL}(\boldsymbol{M})$, $\boldsymbol{T}_{DL}(-\boldsymbol{M})=\boldsymbol{T}_{DL}(\boldsymbol{M})$, where $\boldsymbol{M}$ is the magnetization of ferromagnetic layer[16][17]. In our calculation, the torkances in TM/FM bilayers have been calculated by using Kubo-Bastin linear response formalism as implemented in KKR Green's function method[16].

Specifically, for the self-consistent calculations, a cutoff $\boldsymbol{l}_{max}$=3 had been adopted for the angular momentum expansion, and the Vosko-Wilk-Nussair (VWN) type of local density approximation (LDA) is employed for describing exchange-correction potential[19]. After obtaining self-consistent potentials, the electric field induced torkances are calculated by using $10^6$ $\boldsymbol{k}$-points in Brillouin Zone and 16 energy points by considering phonon scattering at 300 K described by alloy analogy model based on the coherent potential approximation[20].

## Results and discussions

## I. The symmetry imposed torkance tensors in TM/FM bilayers

As shown in Fig.1, we setup the TM/FM bilayers by following the bcc(001) crystal structure with vacuum being simulated by empty spheres. The magnetization is set to be along $\boldsymbol{z}$ direction (perpendicular to the film plane). The bilayers belong to primitive tetragonal crystal structure, therefore there are four symmetries imposed nonzero torkance elements $\boldsymbol{t}_{xx}=\boldsymbol{t}_{yy}$, $\boldsymbol{t}_{xy}=-\boldsymbol{t}_{yx}$. It is easy to identify that the torkances $\boldsymbol{t}_{xx}$ and $\boldsymbol{t}_{yx}$ are corresponding to field and damping like torque, respectively. For instance, by performing mirror operation $M_y$ on the bilayers, since the spin torque and magnetization are axial vectors, while the electric field is polar vector, it follows $\boldsymbol{t}_{xx}(\boldsymbol{M})=-\boldsymbol{t}_{xx}(-\boldsymbol{M})$; $\boldsymbol{t}_{yx}(\boldsymbol{M})=\boldsymbol{t}_{yx}(-\boldsymbol{M})$, which manifest that $\boldsymbol{t}_{xx}$ is the field like (T-odd) torkance and $\boldsymbol{t}_{yx}$ is damping like (T-even) torkance. Since the damping like (DL) torque is decisive for the current induced magnetization switching and dynamics, we are focusing on DL torque in the following discussions.

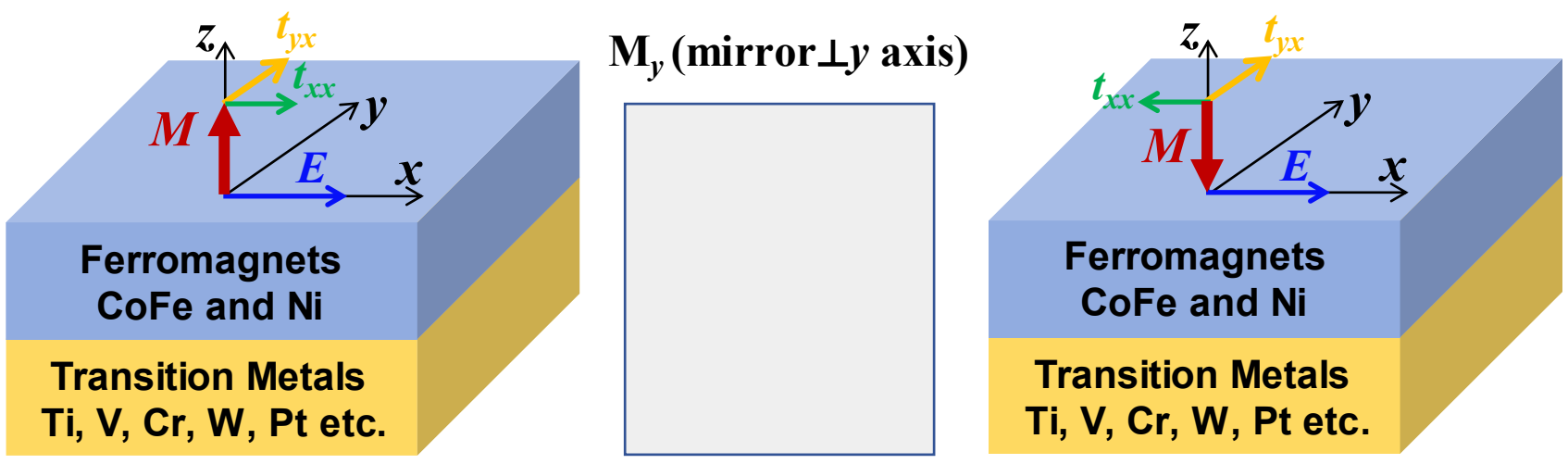

Fig.1 (Left) The schematic diagram of electric field induced spin torque in TM/FM bilayers and Cartesian coordinate system adopted in our calculation. The directions of magnetization $\boldsymbol{M}$, electric field $\boldsymbol{E}$ and torkance $\boldsymbol{t}_{xx}$, $\boldsymbol{t}_{yx}$ are indicated in red, blue, green and yellow arrows respectively. (Right) The symmetry of spin torque in TM/FM bilayers after mirror operation $M_y$.

Based on the calculated torkance, one can evaluate the experimentally measurable current induced damping like effective field $\boldsymbol{B}_{\mathrm{DL}}/\boldsymbol{J}_{\mathrm{c}}$ and torque efficiency $\xi_{\mathrm{DL}}$ in TM/FM bilayers. The torque on FM layer can be written as $\boldsymbol{T}=\boldsymbol{m}\times\boldsymbol{B}_{\mathrm{DL}}$, where $\boldsymbol{m}$ is the total magnetic moment of the ferromagnet layer. Then the effective magnetic field can be expressed as $\boldsymbol{B}_{\mathrm{DL}}=\boldsymbol{T}/\boldsymbol{m}=\boldsymbol{t}_{\mathrm{DL}}\boldsymbol{E}/\boldsymbol{m}$, where $\boldsymbol{t}_{\mathrm{DL}}$ is the DL torkance and $\boldsymbol{E}$ is the electric field. By further considering the charge current density $\boldsymbol{J}_{\mathrm{c}}=\sigma\boldsymbol{E}$, where $\sigma$ is the conductivity of bilayers, one obtains $\boldsymbol{B}_{\mathrm{DL}}/\boldsymbol{J}_{\mathrm{c}}=\boldsymbol{t}_{\mathrm{DL}}/(\boldsymbol{m}\sigma)$. The dimensionless spin-torque efficiency can be further evaluated from the current induced effective field as follows[1]:

$$\xi_{\mathrm{DL}}=\frac{\boldsymbol{B}_{\mathrm{DL}}}{\boldsymbol{J}_c}\boldsymbol{M}_s t_F\frac{2e}{\hbar}=\frac{\boldsymbol{B}_{\mathrm{DL}}}{\boldsymbol{J}_c}\frac{\boldsymbol{m}}{A}\frac{2e}{\hbar}=\frac{\boldsymbol{t}_{\mathrm{DL}}}{\sigma A}\frac{2e}{\hbar}$$

where $\boldsymbol{M}_{\mathrm{s}}$ is the saturation magnetization and $t_{\mathrm{F}}$ is the thickness of ferromagnet layer, $A$ is the interface area of unit cell.

## II. The spin torque in TM/CoFe bilayers

We first calculate the spin torque in TM(6 MLs)/$Co_{50}Fe_{50}$(3 MLs) bilayers, which are directly relevant for SOT-MRAM applications with TM being chosen to be 5*d* heavy metals W, Pt, Au and 3*d* light metals Ti, V, Cr, Cu. By using *s*-*d* model calculations, Kotani *et al*. have found that for transition metals with less than half filling *d* electrons the SHCs are negative, and for more than half filling the SHCs are positive, while the OHCs are always positive. Such physics picture holds quite well and has been confirmed by recent first-principles calculations (for example Fig.2 (b)). As shown in Fig.2 (a), the calculated damping like torkances for TM(Ti, V, Cr, W)/CoFe bilayers are negative, while for TM(Cu, Pt, Au)/CoFe bilayers are positive, which are fully consistent with the sign of SHCs, regardless that their OHCs are all positive. Moreover, the magnitude of torkances and torque efficiency for TM/CoFe bilayers generally follows the trends of SHCs but not OHCs. Particularly, the spin torque efficiency in Ti/CoFe and V/CoFe are found to be vanishing small (negative value at the order of $10^{-4}$), even though Ti and V have very large OHCs (>4000 S/cm). The sizable torkance $\boldsymbol{t}_{DL}$, torque efficiency $\xi_{DL}$ and current induced magnetic field $\boldsymbol{B}_{DL}/\boldsymbol{J}_c$(see supplementary Note 1) in TM/CoFe bilayers only present for 5*d* heavy metals with large spin-orbit coupling and SHCs. Our results confirm that in TM/CoFe bilayers, the spin torque on CoFe can be well and solely attributed to spin Hall mechanism, while the orbital Hall effect in TM is ineffective.

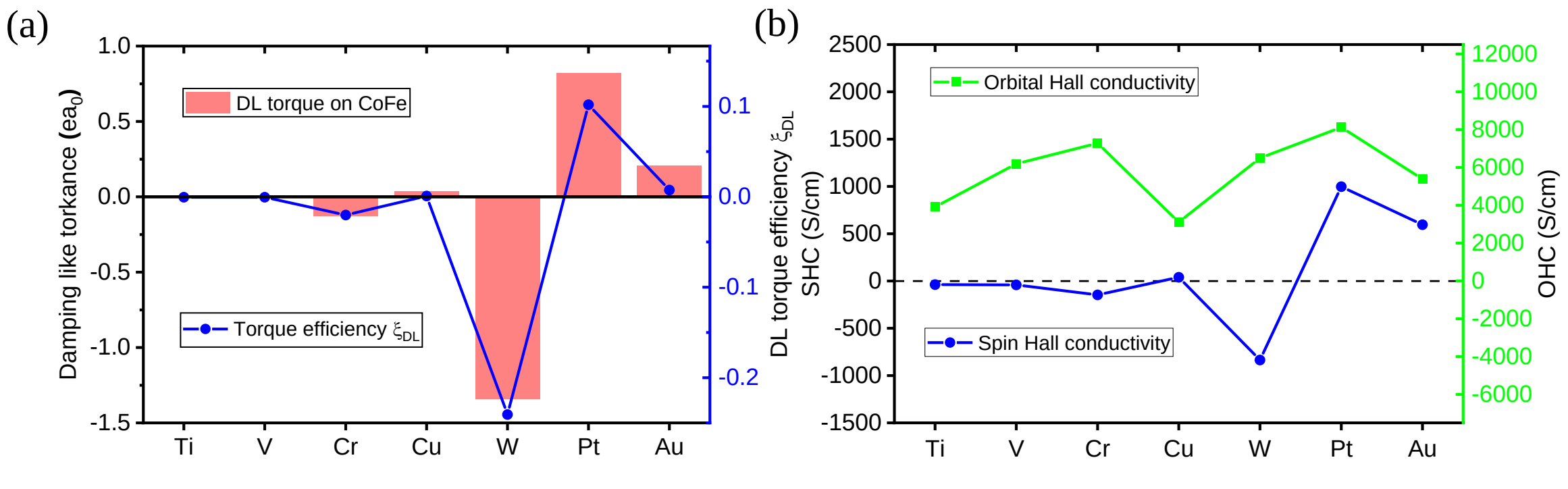

Fig.2 (a) The calculated damping like torkance (red columns refer to left axis) and torque efficiency $\xi_{DL}$ (blue circles refer to right axis) in TM(6 MLs)/CoFe bilayers with TM=Ti, V, Cr, Cu, W, Pt, Au. The atomic unit of torkance is $ea_0$, where $e$ is elementary charge and $a_0$ is Bohr radius. (b) The calculated spin Hall (blue circles refer to left axis) and orbital Hall conductivity (green squares, taken from reference[9]) for bulk transition metals.

To further reveal the spin torque mechanism in TM/CoFe bilayers, the layer resolved torkance and TM thickness dependence of spin torque efficiency $\xi_{DL}$ for TM=Pt, W, Ti, V have been shown in Fig.3. It is clear that the remarkable spin torques for Pt/CoFe and W/CoFe bilayers originate from the large torkance at the interfacial CoFe layer and they show positive and negative interface torkance for Pt and W respectively. This result accords with the picture that the spin current from TMs is absorbed at the TM/CoFe interface and the resultant nonequilibrium spin density exerts torque on FM *via* exchange interaction.

The TM thickness dependence of spin torque is also in line with the spin Hall mechanism. For heavy metals Pt and W with large SHCs, by considering finite spin diffusion length in TMs due to scattering (phonon scattering in our case), the spin torque efficiency originating from spin Hall effect in TMs can be expressed as $\xi_{DL}=\xi_0+\xi_{SH}[1\text{-sech}(d_N/l_{sf})]$, where $\xi_0$ is the interface spin torque contribution, $\xi_{SH}$, $d_N$ and $l_{sf}$ are the spin Hall coefficient, film thickness and spin diffusion length of TMs[21]. The DL torque efficiency $\xi_{DL}$ should first increase with increasing TM thickness and then saturate at thick TM limit, which reflects the bulk spin Hall effect in TMs, as shown in Fig.3 (b) for (W, Pt, Ti,V)/CoFe bilayers.

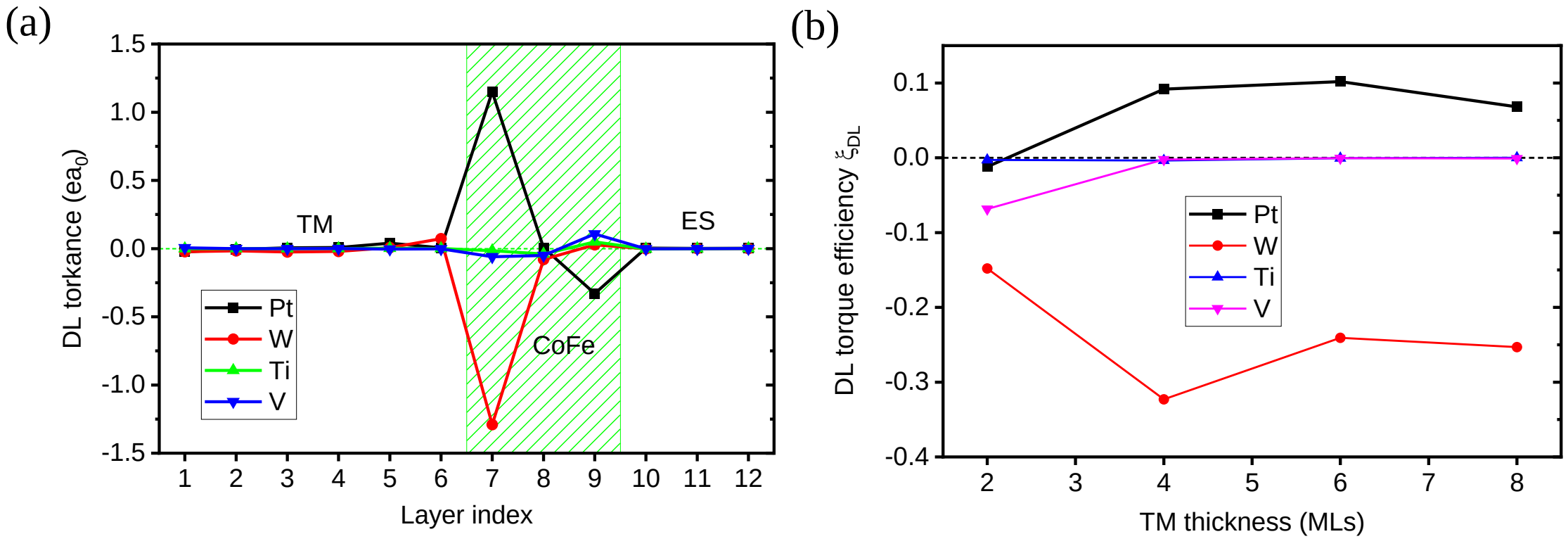

Fig.3 (a) The layer resolved torkance for TM(6 MLs)/CoFe(3 MLs) bilayers with TM=Pt, W, Ti, V. ES are the empty spheres for modeling vacuum at CoFe surface. (b) The DL torque efficiency $\xi_{DL}$ versus TM thickness for TM(*d* MLs)/CoFe(3 MLs) bilayers with *d*=2, 4, 6, 8 MLs. The horizontal dash lines indicate the position of zero torkance and efficiency.

## III. The spin torque in TM/Ni bilayers

The TM/Ni bilayers are supposed to may have orbital Hall contribution from TM on spin torque, due to larger orbital-to-spin conversion ratio in Ni than CoFe[11][12]. When both electric field induced spin and orbital currents in TM are taken into account, the torque efficiency on FM layers is phenomenologically expressed as $\xi_{DL}$~$\sigma_{SH}T_S$+$C_{FM}$·$\sigma_{OH}T_O$, where $\sigma_{SH}$ and $\sigma_{OH}$ are spin and orbital Hall conductivity of TM, $T_S$ and $T_O$ are the spin and orbital current interface transparency[11][12], and $C_{FM}$ is the orbital-to-spin conversion ratio, proportional to the SOC strength in ferromagnets[12]. Such expression also implies the remarkable orbital contribution and self-torque in FM should be present simultaneously. For the studied TM/Ni bilayers, we investigate TMs with various combinations of SHC and OHC. Particularly, the 3*d* light metals Ti and V have small SHC and OHC of opposite signs, 5*d* heavy metal W has large negative SHC and positive OHC, and Pt has large positive SHC and OHC of the same sign.

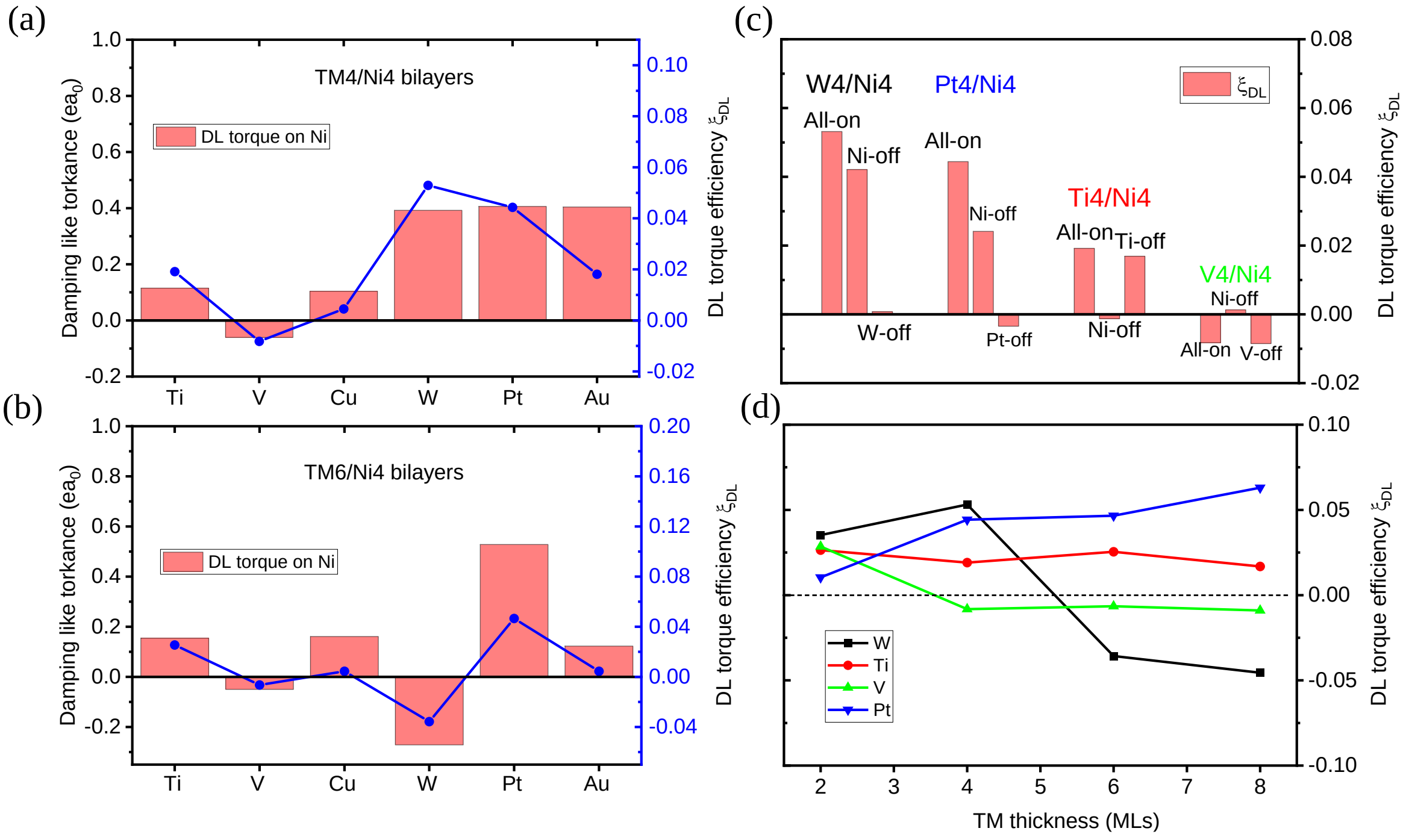

Fig.4 (a), (b) The calculated damping like torkance (red columns refer to left axis) and torque efficiency (blue circles refer to right axis) in TM(4 MLs)/Ni(4 MLs) and TM(6 MLs)/Ni(4 MLs) bilayers with TM=Ti, V, Cu, W, Pt, Au. (c) The DL torkance for TM(4 MLs)/Ni(4 MLs) bilayers with TM=Ti, V, W, Pt when SOC has been turned on for all atoms (All-on), switched off only on TM layer (TM-off) and switched off only on Ni layer (Ni-off). (d) The DL torque efficiency $\xi_{DL}$ versus thickness in TM($d$ MLs)/Ni(4 MLs) bilayers with $d$=2, 4, 6, 8 MLs.

Indeed, the calculated DL torkance and torque efficiency for TM(4, 6 MLs)/Ni(4 MLs) bilayers (hereafter we abbreviate the bilayers as TM4/Ni4 and TM6/Ni4 respectively) shown in Fig.4(a) and(b) are different from the torques in TM/CoFe bilayers. First, the positive torque efficiency $\xi_{DL}$ around 0.02 appears in Ti/Ni4 bilayer, while in the case of Ti/CoFe bilayer, the torque efficiency $\xi_{DL}$ is negative and at the order of $10^{-4}$. Second, the spin torque is positive in W4/Ni4 bilayer, in contrast to the negative torque in W/CoFe bilayer.

An appropriate approach to verify the possible origin of unusual spin torque in TM/Ni bilayers is to adjust the SOC strength on TM and Ni layers when the spin torque is calculated (See supplementary Note 2 for the details on the calculation of spin torque by modulating SOC strength). For instance, when SOC is turned off in transition metals and switched on in Ni layer, there will be no spin but orbital current contribution from TMs on the spin torque, since the orbital Hall effect can be present even without SOC, while the spin Hall effect is vanishing[7]. In this case, the spin torque on Ni may be attributed to the orbital Hall current in TMs and the self spin torque in Ni (please note that the spin Hall conductivity in bulk Ni is the largest among the 3$d$ transition metals). It's worth noting that the orbital Hall contribution should be always accompanied by noteworthy self spin torque in Ni, since both mechanisms rely on strong SOC in ferromagnet. In addition, the presence of strong SOC in heavy metal TM layers may enhance the orbital current transport[22] and its contribution on the spin torque.

One can qualitatively attribute the different contributions to the spin torque by turning on and off the SOC strength on TM and Ni layers according to three different cases: (i).All-on case (SOC on all layers is switched on): Spin torque=TM(SHE)+$TM^{soc}$(OHE)+Ni(self-torque); (ii).Ni-off case (SOC is switched off in Ni, but turned on in TM layers): Spin torque=TM(SHE); (iii).TM-off case (SOC is switched off in TM, but turned on in Ni layers): Spin torque=$TM^{nosoc}$(OHE)+Ni(self-torque). Please note that we have distinguished the orbital Hall contributions as $TM^{soc}$(OHE) and $TM^{soc}$(OHE) when SOC is turned on and off in TM layers.

We first begin the discussions on 5*d* TM4/Ni4 bilayers. For W4/Ni4 bilayer, as shown in Fig.4 (c), when SOC is turned on in Ni but switched off in W layers, the DL torque efficiency $\xi_{DL}$ has been greatly suppressed from around 0.05 to $10^{-4}$, which indicates the SOC driven spin Hall mechanism is responsible for the positive spin torque in W4/Ni4 bilayer. Similarly, for Pt4/Ni4 bilayer, when the SOC is switched off in the Pt layer, the torque efficiency $\xi_{DL}$ has been reduced to a small negative value. To further reveal the origin of spin torque, we calculate the self-torque in Ni layer and qualitatively attribute the sign of spin torque according to different mechanisms (please see supplementary Note 3 and 4 for detailed method and result). The results indicate that in 5*d* TM4/Ni4 bilayers the spin torques induced by orbital Hall effect in TM and self-torque of Ni have opposite sign and nearly canceled each other. Consequently, it exhibits crucial role of SOC driven spin Hall effect for remarkable spin torque in 5*d* TM4/Ni4 bilayers.

We now turn to the case of light 3*d* TM4/Ni4 bilayers. When SOC in TM has been switched off, the damping like torque efficiency $\xi_{DL}$ almost does not change, which is consistent with the vanishing small spin Hall effect in Ti and V. We found that the orbital Hall contribution is always accompanied by noteworthy opposite self spin torque in Ni, which leads to small damping like torque efficiency around 0.02 and -0.01 in Ti4/Ni4 and V4/Ni4 bilayers respectively, despite very large positive orbital Hall conductivity in Ti and V.

Moreover, we examine the spin torque in TM/Ni4 bilayers with various TM layers. As shown in Fig.4 (b), the sign of damping like torkances for 5*d* TM6/Ni4 bilayers aligns with SHCs (Fig.2 (b)), rather than OHCs. This further corroborates the previous conclusions, despite the existence of the orbital Hall effect in 5*d* heavy transition metals, its contribution to spin torque can not compete with the spin Hall contribution. The Ti and V thickness dependence of spin torques in Ti, V(*d* MLs)/Ni(4 MLs) bilayers with *d*=2,4,6,8 MLs have been shown in Fig.4 (d). For Ti, V(*d* MLs)/Ni4 bilayers, the damping like torque efficiency $\xi_{DL}$ on Ni saturates rapidly with the TM thickness increasing from 2 to 8 MLs, which indicates very short orbital current diffusion length (less than 2 MLs) in light 3*d* transition metals[22][23]. On the other hand, for W, Pt(*d* MLs)/Ni4 bilayers, the $\xi_{DL}$ continues increasing with W, Pt thickness, which is an expected result from the bulk spin Hall mechanism in W, Pt. The positive $\xi_{DL}$ in W2/Ni4 and W4/Ni4 bilayers with thin W layer reflect the relatively strong layer thickness dependence of SHE in W films (the obvious thickness dependence of spin torque with thin W layer in W/CoFe bilayers is also evident in Fig.3(b)). In thicker W layers, the SHC will approach the bulk negative value and leads to the negative spin torque in W/Ni4 bilayers.

Indeed, recent experiments demonstrate that the absence of orbit current contribution to spin torque in Ta/FM(Ni, NiFe, CoFeB) bilayers regardless the choice of ferromagnetic layers. The contribution of spin torque signal in Ni itself in Ta/Ni bilayers is also found to be significant[23]. Essentially, there are three possible reasons for the inefficiency of orbital Hall effect on the spin torque in TM/FM bilayers. First, the orbital diffusion length in light metals is very short in comparison with spin diffusion length[22][23][24], and in consequence, very limited orbital current may be transported through the TM/FM interface. Second, even if the orbital current has been transported into FM layer in TM/FM bilayers, it does not directly interact with local magnetization and the generation of spin torque to FM layer relies on the SOC in FM layer, which is a second and indirect effect. Third, the spin torques induced by orbital Hall and self-torque of Ni have opposite sign and nearly canceled each other, which makes the orbital Hall effect being largely suppressed.

## Summary

In summary, we perform first-principles calculations in TM/CoFe and TM/Ni bilayers by elucidating the spin Hall and orbital Hall contribution in transition metals on spin torque. Our investigations reveal that the orbital Hall effect in transition metals does effectively contribute to spin torque. In TM/CoFe bilayers, the damping like spin torque and spin torque efficiency is completely consistent with spin Hall mechanism, while the orbital Hall effect in TM is negligible. In TM/Ni bilayers, spin torques deviating from the orbital Hall mechanism do exist in 3*d* light metal bilayers, while the orbital Hall mechanism compete with self-torque of Ni, remaining small orbital torque efficiency in TM/Ni systems. Our work may shed light on the improvement of spin torque in TM/FM bilayers by focusing and optimizing the spin current mechanism in TM/FM bilayers. For instance, the nonrelativistic spin Hall effect with large spin Hall conductivity and spin Hall angle at room temperature in anisotropic magnetic materials should be promising for spin torque switching of magnetization[25][26].

## Acknowledgement

This work was supported by the National Natural Science Foundation of China (grant No. T2394475, T2394470, 12174129).

## Reference:

[1]. A. Manchon, J. Zelezny, I. M. Miron, T. Jungwirth, J. Sinova, A. Thiaville, K.Garello, and P. Gambardella, "Current-induced spin-orbit torques in ferromagnetic and antiferromagnetic systems", Rev. Mod. Phys. 91, 035004 (2019).

[2]. L. Liu, O. J. Lee, T. J. Gudmundsen, D. C. Ralph, and R. A. Buhrman, "Current-induced switching of perpendicularly magnetized magnetic layers using spin torque from the spin Hall effect", Phys. Rev. Lett. 109, 096602 (2012).

[3]. L. Liu, C.-F. Pai, Y. Li, H. W. Tseng, D. C. Ralph, and R. A. Buhrman, "Spin-torque switching with the giant spin Hall effect of Tantalum", Science 336, 555 (2012).

[4]. A. Manchon, and S. Zhang, "Theory of nonequilibrium intrinsic spin torque in a single nanomagnet", Phys. Rev. B 78, 212405 (2008).

[5]. V. M.Edelstein, "Spin Polarization of conduction electrons induced by electric current in two-dimensional asymmetric electron systems", Solid State Commun. 73, 233(1990).

[6]. I. M. Miron, K. Garello, G. Gaudin, P.-J. Zermatten, M. V. Costache, S. Auffret, S. Bandiera, B. Rodmacq, A. Schuhl, and P. Gambardella, "Perpendicular switching of a single ferromagnetic layer induced by in-plane current injection", Nature (London) 476, 189 (2011).

[7]. H. Kontani, T. Tanaka, D.S. Hirashima, K.Yamada, and J. Inoue, "Giant Orbital Hall effect in Transition Metals: Origin of Large Spin and Anomalous Hall effects", Phys. Rev. Lett. 102, 016601 (2009).

[8]. D. Go, H.-W. Lee, P.M. Oppeneer, S. Blügel, and Y. Mokrousov, "First-principles calculation of orbital Hall effect by Wannier interpolation: Role of orbital dependence of the anomalous position", Phys. Rev. B 109, 174435 (2024).

[9]. S. Mankovsky and H. Ebert, "Spin and orbital Hall effect in nonmagnetic transition metals: Extrinsic versus intrinsic contributions", Phys. Rev. B 110, 184417 (2024).

[10]. D. Go, and H.-W. Lee, “Orbital torque: Torque generation by orbital current injection”, Phys. Rev. Res. 2,013177 (2020).
[11]. D. Go, F.Freimuth, J.-P. Hanke, F.Xue, O.Gomonay, K.-J. Lee, S. Blügel, P.M. Haney, H.-W. Lee, and Y. Mokrousov, “Theory of current-induced angular momentum transfer dynamics in spin-orbit coupled systems”, Phys. Rev. Res. 2, 033401 (2020).
[12]. D. Lee, D. Go, H.-J. Park et al., “Orbital torque in magnetic bilayers”, Nat. Commun. 12, 6710 (2021).
[13]. H. Ebert, D. Ködderitzsch, and J. Minár, “Calculating condensed matter properties using the KKR-Green's function method—recent developments and applications”, Rep. Prog. Phys. 74, 096501 (2011).
[14]. H. Ebert et al., The Munich SPR-KKR Package, version 8.6, http://olymp.cup.uni-muenchen.de/ak/ebert/sprkkr (2017).
[15]. P. M. Haney, and M. D. Stiles,“Current-Induced Torques in the Presence of Spin-Orbit Coupling,” Phys. Rev. Lett. 105, 126602(2010).
[16]. S. Wimmer, K. Chadova, M.Seemann, D. Kodderitzsch, and H. Ebert, “Full relativistic description of spin-orbit torques by means of linear response theory”, Phys. Rev. B 94, 054415 (2016).
[17]. F. Freimuth, S. Blügel, and Y. Mokrousov, “Spin-orbit torques in Co/Pt(111) and Mn/W(001) magnetic bilayers from first-principles”, Phys. Rev. B 90, 174423 (2014).
[18]. K. D. Belashchenko, A. A. Kovalev, and M. van Schilfgaarde, “First-principles calculation of spin-orbit torque in a Co/Pt bilayer”, Phys. Rev. Mater. 3, 011401 (2019).
[19]. S. H. Vosko, L. Wilk, and M. Nusair, “Accurate spin-dependent electron liquid correlation energies for local spin density calculations: a critical analysis”, Can. J. Phys. 58, 1200 (1980).
[20]. H. Ebert, S. Mankovsky, K. Chadova, S. Polesya, J. Minǎr, and D. Ködderitzsch, “Calculating linear-response functions for finite temperatures on the basis of the alloy analogy model”, Phys. Rev. B 91, 165132 (2015).
[21]. K. D.Belashchenko, A. A. Kovalev, and M. van Schilfgaarde, “Interfacial contributions to spin-orbit torque and magnetoresistace in ferromagnet/heavy-metal bilayers”, Phys. Rev. Mater. 3, 011401 (2019).
[22]. Max Rang and Paul J. Kelly, “Orbital relaxation length from first-principles scattering calculations”, Phys. Rev. B 109, 214427 (2024).
[23]. Qianbiao Liu, Lijun Zhu, “Absence of orbital current torque in Ta/ferromagnet bilayers”, arxiv.2501.10260.
[24]. X. Ning, A. Pezo, K.-W. Kim, W. Zhao, K.-J. Lee, and A. Manchon, “Orbital Diffusion, Polarization, and Swapping in Centrosymmetric Metals”, Phys. Rev. Lett. 134, 026303 (2025).
[25]. K. D. Belashchenko, “Exchange-driven spin Hall effect in anisotropic ferromagnets”, Phys. Rev. B 109, 54409 (2024).
[26]. Fanxing Zheng, Jianting Dong, Yizhuo Song, Meng Zhu, Xinlu Li, and Jia Zhang, “Spin Hall effect in 3d ferromagnetic metals for field-free switching of perpendicular magnetization: A first-principles investigation”, Appl. Phys. Lett.126, 9 (2025).